\documentclass[prl,aps,twocolumn,preprintnumbers,superscriptaddress,showpacs]{revtex4-1}

\usepackage{amsmath}
\usepackage{graphicx}
\usepackage{amssymb}
\usepackage{color}

\begin{document}

\title{Analytic continuation with Pad\'{e} decomposition}
\author{Xing-Jie Han}
\affiliation{Institute of Physics, Chinese Academy of Sciences, Beijing 100190,
China}
\affiliation{University of Chinese Academy of Sciences, Beijing 100049, China}
\author{Hai-Jun Liao}
\affiliation{Institute of Physics, Chinese Academy of Sciences, Beijing 100190,
China}
\affiliation{University of Chinese Academy of Sciences, Beijing 100049, China}
\author{Hai-Dong Xie}
\affiliation{Institute of Physics, Chinese Academy of Sciences, Beijing 100190,
China}
\affiliation{University of Chinese Academy of Sciences, Beijing 100049, China}
\author{Rui-Zhen Huang}
\affiliation{Institute of Physics, Chinese Academy of Sciences, Beijing 100190,
China}
\affiliation{University of Chinese Academy of Sciences, Beijing 100049, China}
\author{Zi Yang Meng}
\affiliation{Institute of Physics, Chinese Academy of Sciences, Beijing 100190,
China}
\affiliation{University of Chinese Academy of Sciences, Beijing 100049, China}
\author{Tao Xiang}
\affiliation{Institute of Physics, Chinese Academy of Sciences, Beijing 100190,
China}
\affiliation{University of Chinese Academy of Sciences, Beijing 100049, China}
\affiliation{Collaborative Innovation Center of Quantum Matter, Beijing
100190, China}

\begin{abstract}
  The ill-posed analytic continuation problem for Green's functions or self-energies can be carried out using the Pad\'e rational polynomial approximation.
  However, to extract accurate results from this approximation, high precision input data of the Matsubara Green function are needed.
  The calculation of the Matsubara Green function generally involves a Matsubara frequency summation, which cannot be evaluated analytically.
  Numerical summation is requisite but it converges slowly with the increase of the Matsubara frequency.
  Here we show that this slow convergence problem can be significantly improved by utilizing the Pad\'{e} decomposition approach to replace the Matsubara frequency summation by a Pad\'{e} frequency summation, and high precision input data can be obtained to successfully perform the Pad\'{e} analytic continuation.
\end{abstract}


\maketitle

The theory of the Matsubara Green function provides a convenient and powerful tool to calculate physical properties at finite temperatures~\cite{Mahan-2000}.
In this theory, a Matsubara correlation function corresponding to a dynamic response function measured by experiments is evaluated at a discrete set of imaginary Matsubara frequencies.
An analytic continuation is then performed from this Matsubara correlation function to extract the dynamic response function in real frequency.
This analytic continuation can be carried out straightforwardly if the Matsubara correlation  function can be expressed in a simple analytic formula.
However, in many cases, the Matsubara correlation functions can be evaluated only approximately.
In these cases, the analytic continuation becomes a numerically ill-posed inverse problem, and remains one of the most challenging issues in computational physics.

Different methods have been proposed to perform analytic continuation according
to the accuracy of the input Matsubara correlation functions.
The data of the Matsubara correlation function generated by quantum Monte Carlo simulations (QMC)\cite{Blankenbecler-1981,Hirsch-1983} contain stochastic noise induced by random sampling.
For this kind of data, various numerical algorithms, such as the maximum entropy~\cite{Jarrell-1996,Dominic-2016,Goulko-2017}, the singular value decomposition~\cite{Creffield-1995,Gunnarsson-2010} and the stochastic analytical continuation~\cite{Beach-2004,Sandvik-2016,Qin-2016}, have been developed and successfully applied to various quantum systems.
However, these methods have difficulties in reproducing fine structures in dynamic response functions.
If, on the other hand, high precision input data are available, another method called the Pad\'{e} analytic continuation,~\cite{Vidberg-1977,Beach-2000,Ostlin-2012,Osolin-2013,Schott-2016} can be used.
In this method, the analytic continuation is carried out by interpolating the input date defined in the Matsubara frequency domain with a rational function that is assumed to represent approximately the Matsubara correlation function.
However, the Pad\'{e} analytic continuation is sensitive to the accuracy of the input data.
A tiny noise in the input may lead to a large error in the final result, and high precision calculation of the Matsubara correlation function is needed in order to use the Pad\'{e} formula~\cite{Beach-2000}.

In the Feynman diagram calculation of quantum field theory, a Matsubara correlation function is often determined by a summation over a set of Matsubara frequencies.
In many cases, this summation cannot be carried out analytically and one has to resort to numerical calculations.
A problem often encountered is that the summation converges very slowly with the Matsubara frequency, rendering high precision data required in the Pad\'{e} analytic continuation difficult to obtain.
It is well known that the summation over Matsubara frequencies is equivalent to the contour integration around the poles of Fermi or Bose distribution functions.
The poor convergence of the summation can be overcome by finding new poles of the
Fermi or Bose distribution functions\cite{Taisuke-2007,Croy-2009,Hu-2010,Jie-2011}.
An efficient approach along this line is the Pad\'{e} decomposition first introduced by Ozaki\cite{Taisuke-2007}, and the poles are called Pad\'{e} frequencies.
In this letter, we apply this approach to solve the slow converging problem encountered in the analytic continuation using the Pad\'e rational polynomial approximation.

Let us assume $K(i \omega_n)$ to be the Matsubara Green function corresponding to a dynamic correlation function $K^R(\omega )$, where $\omega_n$ is the Matsubara frequency, and $\omega$ is the real frequency.
As $K(z)$ is analytic in the whole upper complex plan of $z$ excluding the real axis, the retarded Green function $K^R(\omega )$ can be obtained from $K(i\omega_n)$ by analytic continuation
\begin{equation}
  K^R(\omega ) = K(i \omega_n \rightarrow \omega + i 0^+) ,
\end{equation}
provided that the analytic expression of $K\left( i\omega _{n}\right) $ is known.
The spectral function $A\left( \omega \right)$ is determined by the imaginary part of $K^R(\omega  +i0^+)$ by the formula
\begin{equation}
  A( \omega ) =- \frac{1}{\pi} \mathrm{Im} K^{R}\left( \omega +i0^+ \right) .
\end{equation}

\begin{figure}[tbp]
\includegraphics[width=0.7\columnwidth,clip]{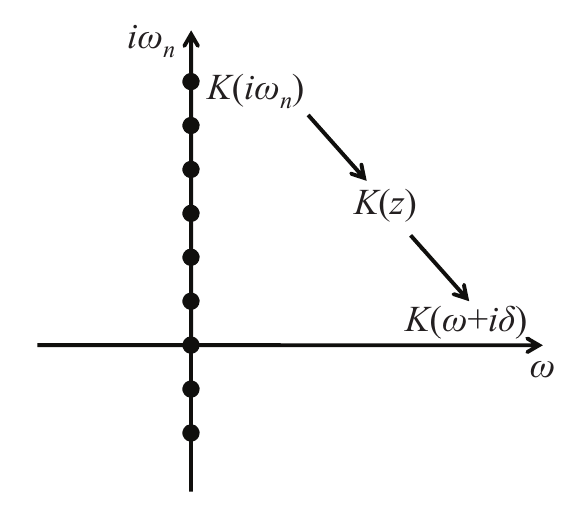}
  \caption{Procedure of the Pad\'e analytic continuation: the Pad\'e rational function $K(z)$, defined by Eq. (\ref{Poly}), is first determined from a set of input data $\{ \omega _{n},K( i \omega _{n})\} $, and then the retarded Green function $K^{R}(\omega +i0^+)$ is obtained from $K(z)$ by substituting $z$ with $\omega +i0^+$. }
\label{fig1}
\end{figure}

However, as already mentioned, the analytic expression of $K\left( i\omega
_{n}\right) $ is not always available, and we can only calculate the correlation
function $K\left( i\omega _{n}\right) $ at some Matsubara frequencies $i\omega _{n}$ numerically. According to a theorem proven by Baym and Mermin~\cite{Baym-1961}, there exists a unique analytic continuation of the Matsubara correlation function, provided the values of $K\left( i\omega _{n}\right) $ for an infinite set of points, including the points at infinity. This is apparently impossible and we have to resort to some
approximations.

Pad\'{e} analytic continuation is based on the assumption that the Matsubara correlation function can be approximately represented by a rational function of degree $r$
\begin{equation}
  K\left( z\right) =\frac{p_{0}+p_{1}z+\cdots +p_{r-1}z^{r-1}}{q_{0}+q_{1}z+\cdots +q_{r-1}z^{r-1}+z^{r}},  \label{Poly}
\end{equation}
where $p_{r}$ and $q_{r}$ are complex coefficients, which can be determined by solving $2r$ linear equations from $2r$ arbitrary but different input points $\left\{ i\omega _{n}, K\left( i\omega _{n}\right) \right\} $~\cite{Beach-2000}.
After the coefficients $\{ p_{r},q_{r} \} $ are determined, we can replace $z$ by $\omega +i 0^+$ to obtain the retarded correlation function $K^R(\omega )$.
The procedure of this analytic continuation is illustrated in Fig.~\ref{fig1}.

\begin{figure}[tbp]
\includegraphics[width=0.85\columnwidth]{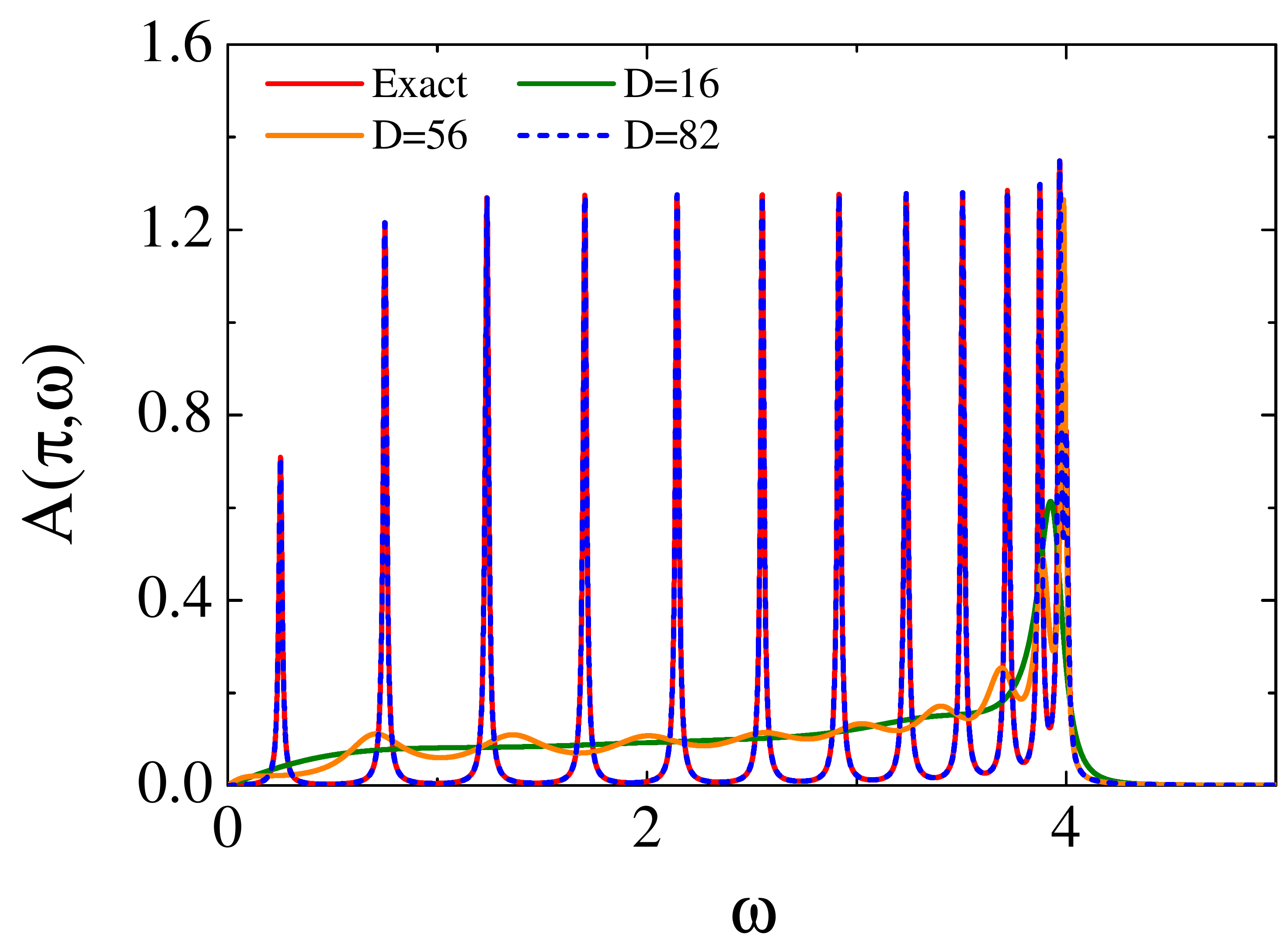}
  \caption{The spectral function of the particle-hole correlation function, $A(\pi, \omega)$, obtained by the Pad\'e analytic continuation with $D=16$ (green solid line), $D=56$ (orange solid line)and $D=82$ (blue dashed line), respectively. The exact result (red solid line) is also shown for comparison.}
\label{fig2}
\end{figure}

When the degree of the rational function $r$ is large, the ratio between the largest and the smallest input imaginary frequencies, $\omega_n$, could be very large.
In this case, high precision calculation is needed to accurately determine the coefficient matrix of the $2r$ linear equations~\cite{Beach-2000}.
To understand this more clearly, let us take the correlation function determined by the particle-hole bubble diagram in one dimension,
\begin{equation}
  K( q,i\omega _{n}) = \frac{1}{\beta N}\sum_{k,i\omega _{m}}G( k,i\omega _{m}) G( q+k,i\omega _{n}+i\omega _{m}) , \label{PH_C}
\end{equation}
as an example to examine how the results of Pad\'{e} analytic continuation are affected by the accuracy of calculations.
In Eq.~(\ref{PH_C}), $G\left( k,i\omega _{m}\right) =1/(i\omega _{m}-\xi _{k})$ is the single-particle Green function with $\xi _{k}=-2\cos k$.
$\omega _{m}$ and $\omega _{n}$ are the fermionic and bosonic Matsubara frequencies.
In the following we use $D$ to denote the decimal digits of the precision, and set $q=\pi $, temperature $T=0.1$, and the lattice size $N=50$.
In the analytic continuation, we replace $z$ by $\omega + i\delta$ with $\delta = 0.01$ in the Pad\'e function (\ref{Poly}).

For this simple correlation function, the Matsubara frequency summation in
Eq.~(\ref{PH_C}) can be carried out exactly.
This gives a rigorous expression
\begin{equation}
K\left( q,i\omega _{n}\right) =\frac{1}{N}\sum_{k}\frac{f\left( \xi
_{k}\right) -f\left( \xi _{k+q}\right) }{i\omega _{n}+\xi _{k}-\xi _{k+q}},
\label{PH_Ex}
\end{equation}
that can be used to benchmark the results obtained numerically with the Pad\'{e} analytic continuation. Here $f\left( x\right)$ in Eq.~(\ref{PH_Ex}) means Fermi distribution function. The analytic continuation of Eq.~(\ref{PH_Ex}) can be carried out by replacing $%
i\omega _{n}$ with $\omega +i\delta $, and the corresponding spectral functions, which would serve as the exact results, can be obtained. For the exact results, we also set $\delta = 0.01$, $T=0.1$ and the lattice size $N=50$.

To perform the Pad\'{e} analytic continuation, we use Eq.~(\ref{PH_Ex}) to generate $2r$ ($r=30$) data points $\left\{ i\omega _{n},K\left( \pi ,i\omega _{n}\right) \right\}$ with precision $D=16$ and $D=56$, respectively.
The Pad\'e coefficients are then determined by solving the $2r$ linear equations numerically.
Fig.~\ref{fig2} compares the spectral function obtained by the Pad\'{e} analytic continuation with the exact result obtained from Eq.~(\ref{PH_Ex}).
With the double precision calculation ($D=16$), we find that the Pad\'e analytic continuation fails to reproduce all the fine structures of the spectral function except the peak structure at the highest frequency.
By increasing the precision to $D=56$, the spectral function calculated by the Pad\'{e} approximation begins to show the multi-peak structures of the spectral function.
However, it still deviates significantly from the exact one. To reproduce accurately the exact result with a difference less than $10^{-15}$, we find that the precision has to be increased to $D=82$.

In the case that the Matsubara frequency summation cannot be evaluated analytically, we have to carry out the summation numerically.
For a direct sum of the Matsubara frequency, one has to introduce a frequency cutoff.
As the summation converges slowly, it is almost impossible to obtain the input data with sufficiently high precision that can be used in the Pad\'e analytic continuation. This problem, however, can be solved by replacing the Matsubara frequency summation with a Pad\'e frequency summation~\cite{Taisuke-2007,Croy-2009,Hu-2010,Jie-2011}.

The Pad\'{e} frequencies are determined by the Pad\'{e} decomposition of the Fermi function
\begin{equation}
  f\left( x\right) =\frac{1}{e^{x}+1}=\frac{1}{2}-\sum_{j=1}^{N\rightarrow \infty }\frac{2\eta _{j}x}{x^{2} + \xi _{j}^{2}},
\end{equation}
where $\pm i\xi _{j}$ and $\eta _{j}$ represent the poles and the corresponding residues of the Fermi function in this decomposition scheme~\cite{Taisuke-2007}.
Table~\ref{PadeTable} lists the values of $\xi _{j}$ and $\eta _{j}$ up to $N=20$.
Unlike the Matsubara decomposition, the poles of the Pad\'{e} decomposition are unevenly distributed.

Using this Pad\'{e} decomposition formula, we can rewrite Eq.~(\ref{PH_C}) as
\begin{eqnarray}
  K\left( q,i\omega _{n}\right) &=&\frac{1}{\beta N}\sum_{k}\sum_{j=-N_P}^{N_P} \eta _{j}G^{\left( 0\right) }\left( k,i\xi _{j}/\beta \right)  \notag \\
  &&G^{\left( 0\right) }\left( q+k,i\omega _{n}+i\xi _{j}/\beta \right) ,
\label{PH_New}
\end{eqnarray}
which differs from Eq.~(\ref{PH_C}) in two respects: (1) the Matsubara frequency summation is changed to the Pad\'{e} frequency summation, and (2) the constant Matsubara residues are replaced by the Pad\'{e} residues.

\begin{table}[tbp]
  \caption{Values of the first 20 pairs of $\xi_j $ and $\eta_j $ defined in the Pad\'{e} spectral decomposition of the Fermi function.}
\label{PadeTable}\centering
\begin{ruledtabular}
  \begin{tabular}{cccccc}
   $j$ &$\eta_j$&$\xi_j/\pi$& $j$ &$\eta_j$&$\xi_j/\pi$\\
    \hline
    1&1.00&1.00&11&1.00&21.0\\
    2&1.00&3.00&12&1.03&23.0\\
    3&1.00&5.00&13&1.22&25.2\\
    4&1.00&7.00&14&1.70&28.1\\
    5&1.00&9.00&15&2.52&32.2\\
    6&1.00&11.00&16&3.90&38.5\\
    7&1.00&13.00&17&6.59&48.7\\
    8&1.00&15.00&18&13.1&67.3\\
    9&1.00&17.00&19&36.8&111\\
    10&1.00&19.00&20&332&332\\

  \end{tabular}
  \end{ruledtabular}
\end{table}

\begin{figure}[tbp]
\includegraphics[width=0.85\columnwidth]{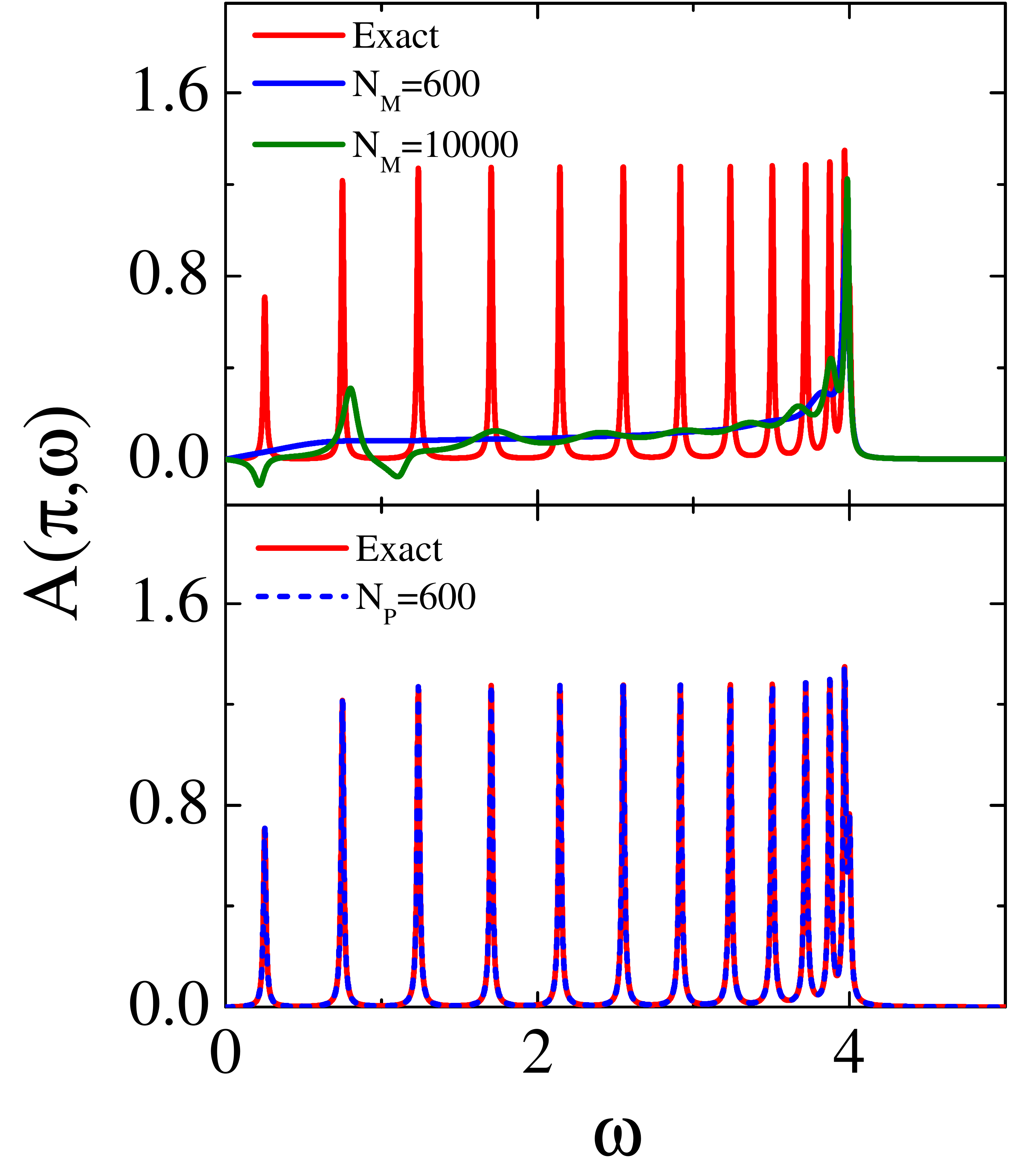}
  \caption{Comparison of the spectral function, $A(\pi, \omega)$, obtained by the Pad\'e analytic continuation with the exact one (red solid lines).
  The input data $\{ i\omega _{n},K(\pi ,i\omega _{n}) \} $ used in the Pad\'{e} analytic continuation are calculated by summing $N_{M}=600$ (blue solid line) and $N_{M}=10000$ (green solid line) Matsubara frequencies (upper panel), and
  $N_{P}=600$ (blue dashed line) Pad\'{e} frequencies (lower panel), respectively.
  }
\label{fig3}
\end{figure}

The Pad\'e decomposition can greatly improve the accuracy in the calculation of the Matsubara frequency summation.
For example, just using $N_P= 70$ Pad\'e frequencies, we can already calculate the value $K( \pi ,i2\pi T) $ with a precision $D=100$.
By contrast, the precision one can obtain for this quantity by directly summing up $N_M = 10000 $ Matsubara frequencies is just $D=3$.

The upper panel of Fig.~\ref{fig3} shows the spectral function of $A\left( \pi ,\omega \right) $ obtained by the Pad\'{e} analytic continuation with the input data calculated by directly summing over $N_{M}=600$ and $N_{M}=10000$ Matsubara frequencies using Eq.~(\ref{PH_C}), respectively.
The exact result is also shown in this figure for comparison.
The difference between the results obtained by the Pad\'e formula and the exact one is apparently very large.
The improvement by increasing $N_M$ from 600 to 10000 is very small.
The spectral function obtained with $N_M=10000$ even becomes negative in some regimes which is clearly unphysical.
For comparison, the lower panel of Fig.~\ref{fig3} shows the spectral function obtained by the Pad\'{e} analytic continuation with the input data calculated by summing over $N_{P}=600$ Pad\'{e} frequencies using Eq.~(\ref{PH_New}), which agrees perfectly with the exact one.

The accuracy of the Pad\'e analytic continuation also depends on the choice of the power $r$ of the rational function.
One should use a sufficiently large $r$ to carry out the analytic continuation so that the results are converged.
As the denominator in the rational function (\ref{Poly}) is a polynomial of order $r$, the number of its roots cannot exceed $r$.
To reproduce accurately a Green's function, $r$ should be equal to or larger than the number of the poles of that function.
However, if $r$ is larger than the number of poles, the presence of the extra poles would also introduce errors in the calculation since it is difficult to make the
contribution of these redundant poles cancel each other.
In the case that the exact number of poles is unknown, it is desired to use a larger $r$ and to carry out the calculation with high precision.
Some other schemes have also been proposed to deal with this so-called pole-zero pairs problem~\cite{Beach-2000,Ostlin-2012,Osolin-2013,Schott-2016}.

In summary, we have shown that high precision input of the Matsubara Green function is needed to carry out the analytic continuation using the Pad\'e rational polynomial approximation.
When the summation over Matsubara frequencies cannot be carried out analytically, the replacement with summation over Pad\'{e} frequencies can significantly improve the convergent behavior.
This approach can be used to calculate the Matsubara Green function with high precision. It can be also used in the calculation of quantum Monte Carlo and dynamic mean-field theory, whenever a Matsubara frequency summation is present.

This work is supported by the National Natural Science Foundation of China (Grants No. 11474331 and No. 11190024)

\end{document}